%% file: main.tex
\documentclass[reprint, amsmath, amssymb, aps, superscriptaddress, pra]{revtex4-1}
\usepackage{format}
\usepackage{soul} 
\usepackage[usenames,dvipsnames]{xcolor}
\allowdisplaybreaks

\usepackage{soul} 
\usepackage{hyperref}
\usepackage{orcidlink}

\begin{document}

\preprint{APS/123-QED}

\title{ Highly elastic collisions of ultracold asymmetric top molecules in a static electric field }

\author{Reuben R. W. Wang\orcidlink{0000-0002-1069-9746}} 
\affiliation{ ITAMP, Center for Astrophysics $|$ Harvard \& Smithsonian, Cambridge, Massachusetts 02138, USA }
\affiliation{ Department of Physics, Harvard University, Cambridge, Massachusetts 02138, USA }

\date{\today} 

\begin{abstract}

We propose that the application of static electric fields to reactive CaNH$_2$ molecules can result in large ratios of elastic to lossy collisions at ultracold temperatures.  
In the first rotationally excited manifold, the asymmetric top structure of CaNH$_2$ results in a $\approx 226$ MHz splitting of the opposite parity $K$-doublet states. 
This modest splitting allows electric fields below a kilovolt per centimeter to polarize the molecules and induce long-range dipole-dipole interactions. 
By increasing the electric field, we find that dipole-dipole attraction can become sufficiently large to induce weakly bound octatomic field-linked states of [CaNH$_2$]$_2$. 
In the absence of these field-linked states, we identify select molecular states where dipolar interactions allow elastic collisions to dominate over lossy collisions by up to three orders of magnitude, while maintaining loss rate coefficients at the $10^{-14}$ cm$^3$s$^{-1}$ level. These results establish that evaporative cooling is a viable route to quantum degenerate gases of laser coolable asymmetric top molecules.   
Access to such low entropy ensembles of asymmetric top molecules holds promise to push the frontiers of quantum simulation, state-resolved chemistry, and searches for beyond the Standard Model physics.

\end{abstract}

\maketitle

\section{ Introduction \label{sec:introduction} }

A generic feature of ultracold polyatomic molecules is the presence of closely spaced, opposite-parity doublets. 
Several appealing applications of these parity doublets have been identified, for instance as synthetic spin-1/2 degrees of freedom for quantum simulation \cite{Robichaud26_Nature, Tao26_preprint}, or as a handle to polarize the molecules for searches of beyond the Standard Model physics \cite{Anderegg23_Science, Kozyryev17_PRL, Jadbabaie23_NJP}.
In asymmetric top molecules, broken cylindrical symmetry results in $K$-doublets: opposite parity superpositions of states with clockwise and counterclockwise rotation about the molecular axis.
These $K$-doublets exist in the vibrational and electronic ground state, offering intrinsically long coherence times in well isolated ultracold environments. 

Laser cooling presents a robust route to achieving dense ultracold samples of polyatomic molecules \cite{Augenbraun20_PRX, Vilas22_Nature, Vilas26_PRX}, recently demonstrated in asymmetric top CaNH$_2$ \cite{Li26_arxiv}. 
A natural next step to bring them further into quantum degenerate temperatures is evaporative cooling, proven successful in diatomic species \cite{Schindewolf22_Nat, Lin26_arxiv, Bigagli24_Nat, Shi26_NatPhys, Jung26_arxiv}. 
For efficient evaporative cooling to occur, a gas taken out of equilibrium must be able to rethermalize through elastic collisions \cite{Wang21_PRA, Wang24_PRR, Wang24_PRA}.     
In dipolar molecular gases, this necessity has been realized through the use of collisional shielding techniques, leveraging externally applied electromagnetic fields to strongly suppress lossy collisions between molecules while simultaneously enhancing elastic collisions \cite{Avdeenkov06_PRA, Lassabliere18_PRL, Karman18_PRL, Karman25_PRXQ}. 

In this work, we show that the application of static electric fields to dipolar CaNH$_2$ can lead to elastic-to-loss collisional rate ratios exceeding three orders of magnitude. 
As larger electric fields are applied, the induced molecular dipole moment  increases, resulting in stronger dipole-dipole interactions and larger elastic cross sections. Simultaneously, rotational van der Waals interactions cause the lower and upper $K$-doublet states to repel, resulting in the upper manifold developing a repulsive barrier against the short-range. These barriers persist even in the presence of electric fields sufficient to polarize the molecules.  
Because loss rate suppression arises from the intrinsic $K$-doublet structure, this collisional stabilization mechanism is expected to be generic to alkaline-earth monoamide molecules, which we also demonstrate with SrNH$_2$.
We therefore propose that this class of molecules can be evaporatively cooled to quantum degeneracy. 

The remainder of this paper is organized as follows. In Sec.~\ref{sec:formulation}, we introduce an effective Hamiltonian model of CaNH$_2$ and analyze its response to a static electric field. Sec.~\ref{sec:collisions} presents the main results of this work, showcasing highly elastic scattering and field-linked bound states in ultracold two-body collisions. Finally, a summary and outlook are provided in Sec.~\ref{sec:summary}.

\section{ Formulation \label{sec:formulation} }

\begin{figure}[ht]
    \centering
    \includegraphics[width=\linewidth]{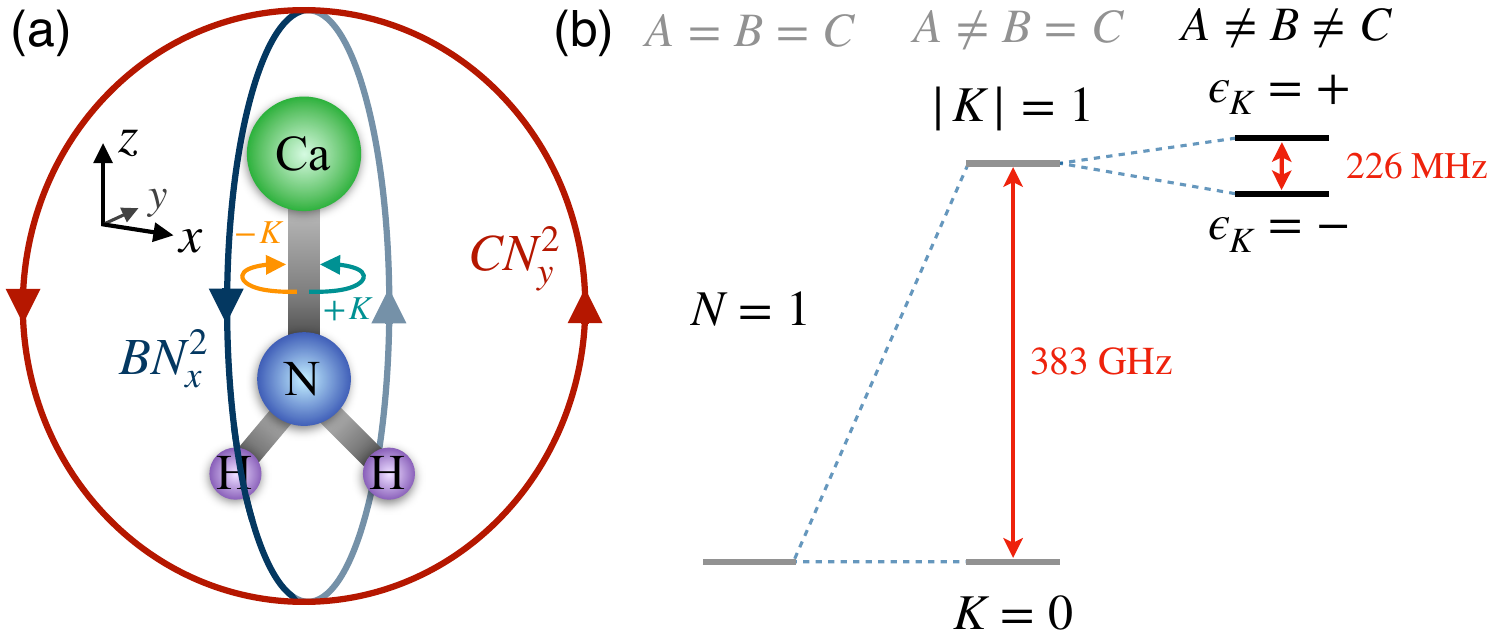}
    \caption{ Molecular rotations of an asymmetric top. (a) Visualization of the principal rotation axes in CaNH$_2$. (b) Rotational structure resulting from the asymmetric top geometry of CaNH$_2$, restricted to the $N=1$ manifold ignoring fine and hyperfine structure. The degeneracy in states of different $|K|$ is lifted when spherical symmetry is broken, while the degeneracy in $K = \pm 1$ states is lifted when cylindrical symmetry along the molecular axis is broken. }
    \label{fig:CaNH2_rotations}
\end{figure}

\begin{table}[ht]
    \centering
    \begin{tabular}{lcc}
        \hline
        constant & symbol & value \\
        \hline
        dipole moment \cite{Marr95_JCP} & $d_0$ & 1.74 D \\
        rotational constants \cite{Brewster00_JCP} & $A$ & 392.127 GHz \\
        & $B$ & 9.0090649 GHz \\
        & $C$ & 8.7827559 GHz \\
        spin-rotation couplings \cite{Brewster00_JCP} & $\epsilon_{xx}$ & 32.063 MHz \\
        & $\epsilon_{yy}$ & 41.110 MHz \\
        & $\epsilon_{zz}$ & 45.7 MHz \\
        \hline
    \end{tabular}
    \caption{ Molecular constants for CaNH$_2$. }
    \label{tab:CaNH2_constants}
\end{table}
  
In its $\tilde{X} ^{2}A_1$ electronic ground state, CaNH$_2$ is an asymmetric top molecule well modeled by the effective Hamiltonian \cite{Augenbraun20_PRX}
\begin{align}
    {\cal H}
    &=
    {\cal H}_{\rm rot}
    +
    {\cal H}_{\rm sr} \nonumber\\
    &=
    ( A N_z^2 + B N_x^2 + C N_y^2 )
    +
    \sum_{\alpha} \epsilon_{\alpha \alpha} N_{\alpha} S_{\alpha}
\end{align}
where $A > B > C$ are rotational constants, $d_0$ is the molecular-frame dipole moment, $\epsilon_{\alpha \alpha}$ are the diagonal spin-rotation coupling constants, and the sum indices run over molecule-fixed axes $\alpha = x, y, z$. The molecular $z$-axis lies along the Ca-N bond. 
We denote lab-fixed coordinates with $(X, Y, Z)$. 
The CaNH$_2$ molecular parameters are provided in  Tab.~\ref{tab:CaNH2_constants}, showcasing its near prolate symmetry as quantified by Ray's asymmetry parameter $\kappa = (2B - A - C)/(A - C) \approx -0.999$. 
Since $B \approx C$, the rotational Hamiltonian is better cast in the form
\begin{align} \label{eq:Hrotation}
    \mathcal{H}_{\rm rot}
    &=
    \frac{1}{2} (B + C) \boldsymbol{N}^2
    +
    \frac{1}{2} \left[ 2 A - (B + C) \right] N_z^2 \nonumber\\
    &\quad 
    +
    \frac{1}{4} (B - C)
    ( N_+^2 + N_-^2 ),
\end{align} 
where $N_{\pm} = N_x \pm i N_y$ are the angular momentum ladder operators. 
The first line of Eq.~(\ref{eq:Hrotation}) is diagonal in the symmetric top basis $|N, K, M\rangle$, while the second line mixes states of $K$ that differ by $\pm 2$. This mixing lifts the degeneracy between $K = \pm 1$ states, a result of the broken cylindrical symmetry along the molecular $z$ axis due to the imbalanced $B$ and $C$ rotational constants, illustrated in Fig.~\ref{fig:CaNH2_rotations}.

The open-shell electronic structure of CaNH$_2$ also causes molecular rotations to significantly couple to the valence electron spin via the electron orbital angular moment. A Van Vleck transformation then produces the effective spin-rotation fine structure Hamiltonian:
\begin{align}
    {\cal H}_{\rm sr} 
    &= 
    \frac{ 1 }{ 4 }
    (\epsilon_{xx} - \epsilon_{yy}) 
    (N_+ S_+ + N_- S_-) \nonumber\\
    &\quad
    +
    \frac{1}{4}
    (\epsilon_{xx} + \epsilon_{yy}) 
    (N_+ S_- + N_- S_+) 
    +
    \epsilon_{z z} N_{z} S_{z},
\end{align}
where this coupling makes relevant the total angular momentum  $\boldsymbol{J} = \boldsymbol{N} + \boldsymbol{S}$ of the molecule. 
Due to the highly ionic calcium-amide bond, the unpaired valence electron is highly polarized and centered almost exclusively on the calcium atom which has zero nuclear spin. As such, we expect the hyperfine couplings to be small. Moreover, the nuclear spins are expected to play only a spectator role in the long-range dynamics of concern here, so we ignore the molecular hyperfine structure in this work.

\subsection{ Restriction to the \textit{N}=1 manifold }

To simplify our analysis, we restrict ourselves to the $N = 1, |K| = 1$ manifold. 
This truncation is warranted as we notice that the coefficients $(B+C)/2$ and $A - (B+C)/2$ all lie in the 10s to 100s of GHz range.   
Meanwhile, the spin-rotation and $K$-doublet (discussed below) structures are 10s to 100s of MHz, establishing a large separation of energy scales for considerations of only the closely degenerate subspace.  
In the reduced subspace, the bare molecular Hamiltonian has matrix elements in the Hund's case (b) symmetric top basis $|J; m_J, K\rangle$ given by 
\begin{subequations}
\begin{align}
    & \bra{ J; m_J, K' }
    \mathcal{H}_{\rm rot}
    \ket{ J; m_J, K } \\
    &=
    \left( A + \frac{B + C}{2} \right)
    \delta_{K',K} \nonumber\\
    &\quad
    +
    \frac{1}{4} (B - C)
    \sqrt{ 2 - K (K \pm 1) } \nonumber\\
    &\quad\quad \times 
    \sqrt{ 2 - (K \pm 1) (K \pm 2) }
    \delta_{K',K \pm 2}, \nonumber\\ 
    & \langle J; m_J, K' |
    {\cal H}_{\rm sr}
    | J; m_J, K \rangle \\
    &=
    f_1(J)
    \left[
    \epsilon_{z z} K^2
    +
    \frac{1}{2}
    (\epsilon_{xx} + \epsilon_{yy}) 
    ( 2 - K^2 )
    \right]
    \delta_{K',K} \nonumber\\
    &\quad 
    +
    \frac{1}{4}
    (\epsilon_{xx} - \epsilon_{yy}) 
    f_1(J) \nonumber\\
    &\quad\quad \times 
    \sqrt{ (1 \mp K) (\mp K) (2 \pm K) (3 \pm K) }
    \delta_{K',K \pm 2}, \nonumber
\end{align} 
\end{subequations}
where $f_1(J) = [J (J + 1)/4 - 11/16]$. See the Supplemental Material for more details on the matrix elements \cite{SI}.

\subsection{ \textit{K}-doublets }

As mentioned, asymmetric top molecules possess a parity doublet due to mixing between the $|K|=1$ states:
\begin{align} \label{eq:parity_eigenstates}
    \ket{ J; m_J, \epsilon_{K} }
    &=
    \frac{ 1 }{ \sqrt{ 2 } }
    \left(
    \ket{ J; m_J, +|K| }
    +
    \epsilon_{K}
    \ket{ J; m_J, -|K| }
    \right),
\end{align}
where $\epsilon_{K} = \pm 1$ labels the parity quantum number. 
These correspond to the conventional $1_{10}$ and $1_{11}$ asymmetric top states in $N_{K_a, K_c}$ notation \cite{Hirota85_Springer}. 
To see the origin of this mixing, we consider fixed $(J, m_J)$ quantum numbers where we can construct a simple $2 \times 2$ matrix representation of the molecular Hamiltonian in the basis $| J; m_J, K \rangle = | J; m_J, -1 \rangle, | J; m_J, +1 \rangle$:
\begin{align}
    \boldsymbol{{\cal H}}_{\rm mol}^{(J, m_J)}
    &=
    E_0(J) \boldsymbol{I}
    +
    \begin{pmatrix}
        0 & W(J) \\
        W(J) & 0
    \end{pmatrix},
\end{align}
where $E_0(J) = f_1(J) [ \epsilon_{zz} + (\epsilon_{xx} + \epsilon_{yy})/2 ]$ and $W(J) = [(B - C) + f_1(J) (\epsilon_{xx} - \epsilon_{yy})]/2$. $\boldsymbol{I}$ is the identity matrix. Diagonalizing this matrix immediately gives the eigenstates in Eq.~(\ref{eq:parity_eigenstates}) with corresponding eigenenergies $E_0 + \epsilon_K W$. 
This pair of opposite parity states is the so-called $K$-doublet. 

A similar parity doublet exists in polyatomic CaOH molecules when its first vibrational bending mode is excited. 
This bending vibration breaks the otherwise linear geometry of CaOH, resulting in an effective asymmetric top structure.  
Coherence times of several hundred milliseconds were observed in the $\ell$-doublet of CaOH \cite{Robichaud26_Nature}, ultimately limited by the radiative lifetime of the vibrational excitation around $0.7$ s \cite{Hallas23_PRL}. 
In contrast, the $K$-doublet in nonlinear asymmetric top molecules exist in the vibrational ground state, making for much longer coherence times. 
In particular, the $K$-doublet states discussed are metastable, because any decay to the $K = 0$ manifold must be accompanied by a triplet to singlet transition in the two hydrogen spins, which is only weakly coupled by hyperfine interactions \cite{Yang26_arxiv}.

\subsection{ Application of a static electric field }

\begin{figure}[ht]
    \centering
    \includegraphics[width=1\linewidth]{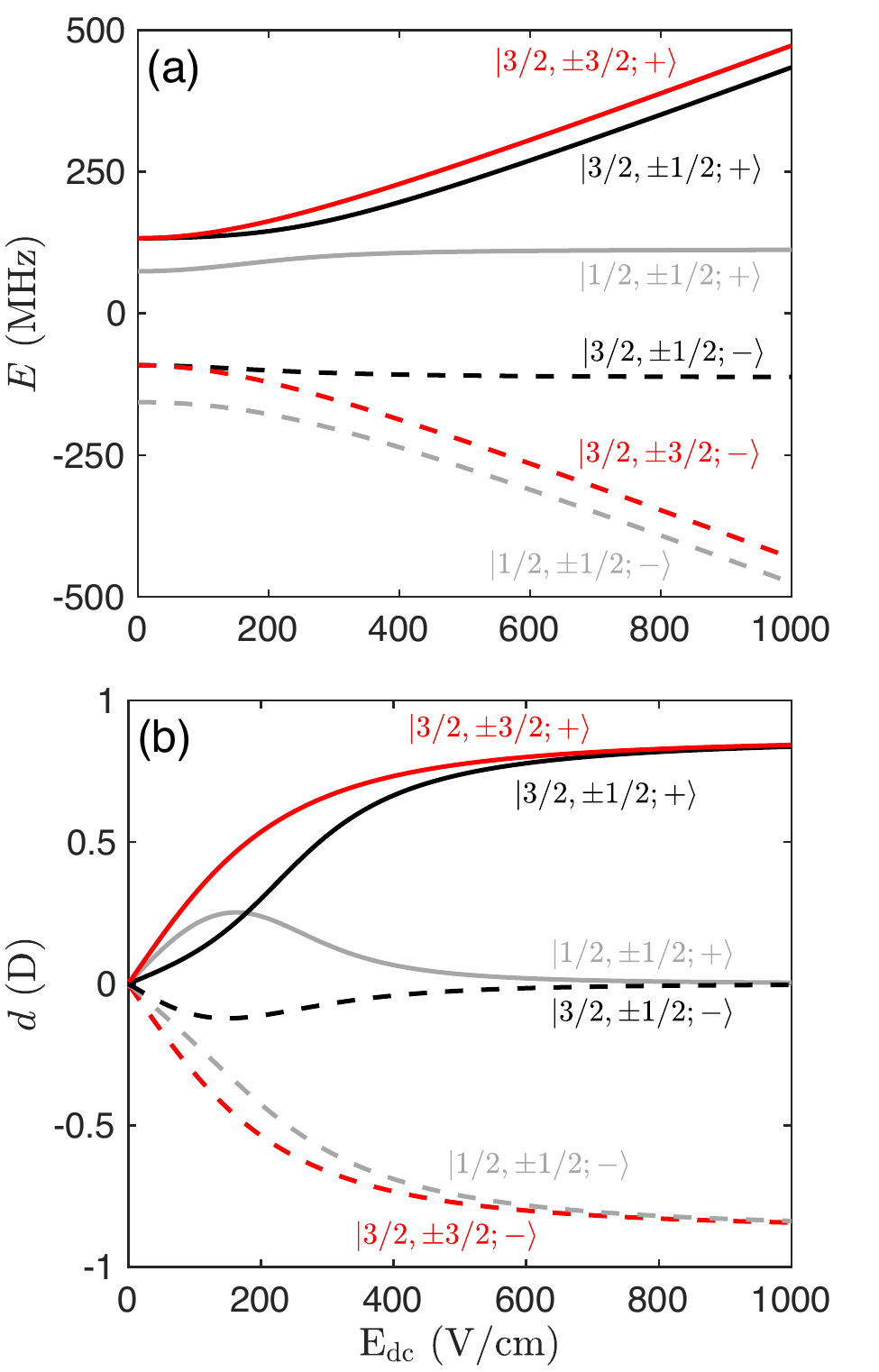}
    \caption{ (a) The molecular spectrum of CaNH$_2$ in the reduced $N=1, |K|=1$ subspace, as a function of electric field. (b) Corresponding induced dipole moments as a function of electric field. In both subplots, the energies and dipole moments correspond to states labeled by the field-dressed quantum numbers $|\tilde{J}, m_J; \tilde{\epsilon}_{K}\rangle$. The $J=3/2$ states with $m_J = \pm 1/2$ ($\pm 3/2$) manifold are colored black (red), while the $J = 1/2$ states are colored gray. Even parity states are plotted as solid curves whereas the odd parity states are dashed. }
    \label{fig:CaNH2_rotationsNdipoles}
\end{figure}

By applying a static electric field $\boldsymbol{{\rm E}}_{\rm dc}$, lab-frame dipole moments can be induced in the molecules, significantly enhancing their long-range interactions. A dipolar molecule couples to the electric field through the interaction ${\cal H}_{\rm dc} = -\boldsymbol{d}_0 \cdot \boldsymbol{{\rm E}}_{\rm dc}$, with matrix elements given by
\begin{align}
    & \langle N', K', M'|
    -\boldsymbol{d}_0 \cdot \boldsymbol{{\rm E}}_{\rm dc}
    | N, K, M \rangle \nonumber\\
    &=
    -d_0 {\rm E}_{\rm dc}
    (-1)^{M' - K'}
    \sqrt{ (2 N' + 1) (2 N + 1) } \nonumber\\
    &\quad\quad \times 
    \begin{pmatrix}
        N' & 1 & N \\
        -K' & 0 & K 
    \end{pmatrix}
    \begin{pmatrix}
        N' & 1 & N \\
        -M' & 0 & M 
    \end{pmatrix},
\end{align}
where $\boldsymbol{d}_0$ is the molecular dipole operator. 
The static electric field will couple states of different total angular momentum $J$ and parity $\epsilon_{K}$, resulting in the field-dressed eigenstates $|\tilde{J}; m_J, \tilde{\epsilon}_{K}\rangle$ that adiabatically correlate the zero-field $|{J}; m_J, {\epsilon}_{K}\rangle$ states.  
We plot the field-dressed molecular spectrum in Fig.~\ref{fig:CaNH2_rotationsNdipoles}(a), and the corresponding induced dipole moments in Fig.~\ref{fig:CaNH2_rotationsNdipoles}(b). 
Although a non-zero lab-frame dipole moment is induced in all states at small electric fields, the high-field limit causes the $|1/2, \pm 1/2; +\rangle$ and $|3/2, \pm 1/2; -\rangle$ states to become field-insensitive and thus without lab-frame dipole moments. 
Molecules prepared in these dipole-free states are therefore not ideal to realize strongly interacting molecular gases at high fields. 

\section{ Ultracold two-body collisions \label{sec:collisions} }

Having formulated the relevant dressed molecular structure, we now turn to the ultracold collisions of CaNH$_2$ dipolar molecules. 
We consider that the molecules interact at long-range predominantly via dipole-dipole interactions:
\begin{align}
    V_{\rm dd}(\boldsymbol{r})
    &=
    \frac{ \boldsymbol{d}_{0,A} \cdot \boldsymbol{d}_{0,B} - 3 ( \boldsymbol{d}_{0,A} \cdot \hat{\boldsymbol{r}}) ( \boldsymbol{d}_{0,B} \cdot \hat{\boldsymbol{r}}) }{ 4\pi \epsilon_0 r^3 }, 
\end{align}
where $\boldsymbol{r}$ is the intermolecular coordinate and $\boldsymbol{d}_{0, i}$ are the dipole operators of molecules $i = A, B$. 

With our focus on suppressing collisional loss, we restrict considerations to molecules prepared in the upper $\epsilon_K = +$ manifold.
In this upper manifold, dipole-dipole interactions between two CaNH$_2$ molecules result in repulsive van der Waals interactions that naturally shield the molecules against the short-range even at zero field; an intrinsic property that is otherwise absent in linear or symmetric top rigid-rotor molecules. 
Ignoring electronic contributions, the repulsive van der Waals coefficient is approximated by $C_{6, {\rm rep}} = [d_0^2/(4\pi\epsilon_0)]^2 / [12(B - C)] \approx 5.3 \times 10^5$ au.

Because calculations are restricted to the $N=1$ manifold, short-range van der Waals interactions from other rotational manifolds and electronic dispersion are included through an isotropic $-C_{6, {\rm att}} / r^6$ attractive interaction. 
The electronic contribution to $C_{6, {\rm att}}$ can be estimated with the single-resonance London dispersion formula $C_{6, {\rm att}}^{\rm elec} \approx (3/4) \hbar\omega_0 [\alpha_0/(4\pi\epsilon_0)]^2 \approx 1515$ au, where $\omega_0$ is the $\tilde{X} \rightarrow \tilde{A}$ transition center frequency and $\alpha_0$ is the valence contribution to the static polarizability. 
Meanwhile, the attractive rotational contribution predominantly arises from couplings to the $N=2$ manifold, approximated by $C_{6, {\rm att}}^{\rm rot} \approx [d^2/(4\pi\epsilon_0)]^2 (3/8) / (B + C) \approx 30,460$ au.  
In the absence of an experimentally measured value for $C_{6, {\rm att}}$, we utilize a conservatively large value of $C_{6, {\rm att}} = 5 \times 10^4$ au. 
This conservative value is still an order of magnitude smaller than $C_{6, {\rm rep}}$, with variations to it only slightly changing the reactive rates but bearing little influence on the elastic and inelastic rates. The position of observed resonances are also largely insensitive to $C_{6, {\rm att}}$ at this order of magnitude.

For the fermionic isotopologues of CaNH$_2$ of interest here, the zero-field elastic scattering is unfortunately suppressed at ultracold temperatures as follows from Wigner's threshold law for $p$-wave collisions \cite{Sadeghpour00_JPB}. We now show that applying electric fields to induce first-order dipolar interactions can significantly enhance elastic scattering, while maintaining high short-range shielding barriers to maintain low loss rates.

\begin{figure}[ht]
    \centering
    \includegraphics[width=\linewidth]{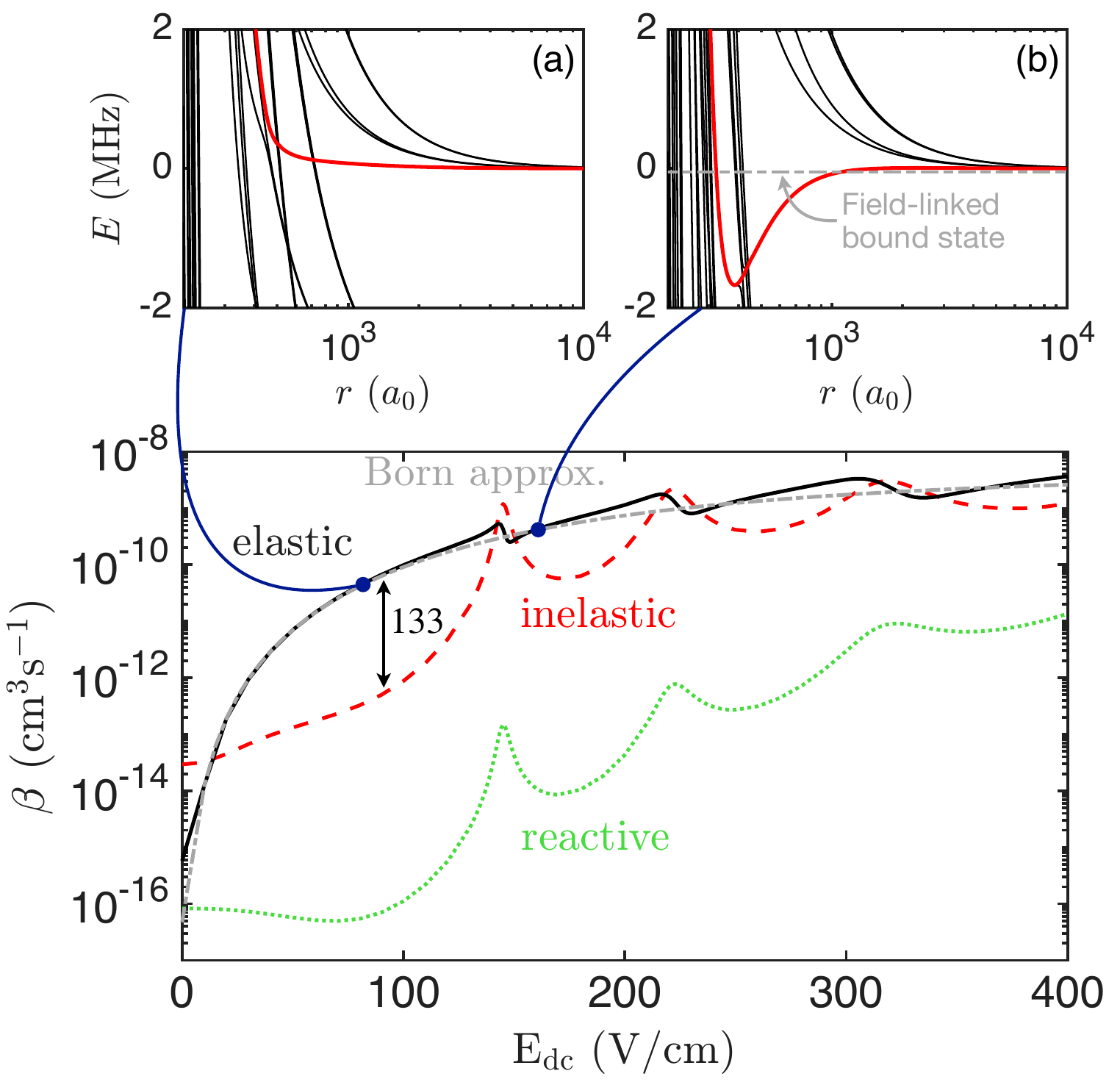}
    \caption{ Upper panels: Close up plots of the potential energy curves that adiabatically connect to the $| 3/2, +3/2; + \rangle$ state at (a) ${\rm E}_{\rm dc} = 80$ V/cm and at (b) ${\rm E}_{\rm dc} = 160$ V/cm. The latter field value results in a field-linked bound state (gray dashed line), existing in the red highlighted potential curve that interpolates across narrow non-adiabatic crossings.
    Lower panel: Elastic (solid black), inelastic (dashed red) and reactive (dotted green) rate coefficients as a function of applied electric field for colliding molecules prepared in the $| \tilde{J}, m_J; \tilde{\epsilon}_K \rangle = | 3/2, +3/2; + \rangle$ state. The elastic rate coefficient obtained from the Born approximation is plotted as the dashed-dotted gray curve. }
    \label{fig:ICS_vs_Edc__JmJ3half}
\end{figure}

\subsection{ Elastic scattering and field-linked states }

At ultracold temperatures, the close-to-threshold elastic scattering behavior can be intuited from the strength of electric field induced dipole moments. In particular, we expect the low energy integral elastic cross section for dipolar scattering to be well described with a simple analytic expression \cite{Bohn09_NJP, Bohn14_PRA}: 
\begin{align} \label{eq:Born_approximation}
    \sigma_{\rm el}
    &\approx 
    \frac{ 32\pi }{ 15 } a_d^2,
\end{align}
derived through the Born approximation.
Evident from Eq.~(\ref{eq:Born_approximation}), the threshold elastic cross section is completely determined by the dipole length $a_d = \mu d^2 / (4 \pi \epsilon_0 \hbar^2)$ with effective dipole moment $d$. 
Therefore, we expect from Fig.~\ref{fig:CaNH2_rotationsNdipoles}(b) that elastic scattering rates for molecules prepared in $J=1/2$ to be large at electric fields of ${\rm E}_{\rm dc} \approx 150$ V/cm. Meanwhile, elastic scattering should increase monotonically with larger electric fields for $J=3/2$ molecules, within the regime where other rotational manifolds are not significantly coupled by the field dressing.

We can verify the approximation above through rigorous numerical close-coupling calculations, that provide us the elastic, inelastic and short-range reactive integral cross sections \cite{Quemener17_RSC}.  
In the current study, scattering calculations are performed with an adaptive step size version of the Manolopoulos log-derivative propagation algorithm \cite{Manolopoulos86_JCP}, using up to $L = 15$ partial waves to converge calculations. 
We apply a universal-loss short-range absorbing boundary condition \cite{Wang15_NJP} to account for chemical reactions of the molecular free radicals at close proximity.   
In turn, these cross sections also provide us the experimentally relevant rate coefficient $\beta = \sigma v_r$, where $v_r$ is the relative collision velocity.       

\begin{figure}[ht]
    \centering
    \includegraphics[width=\linewidth]{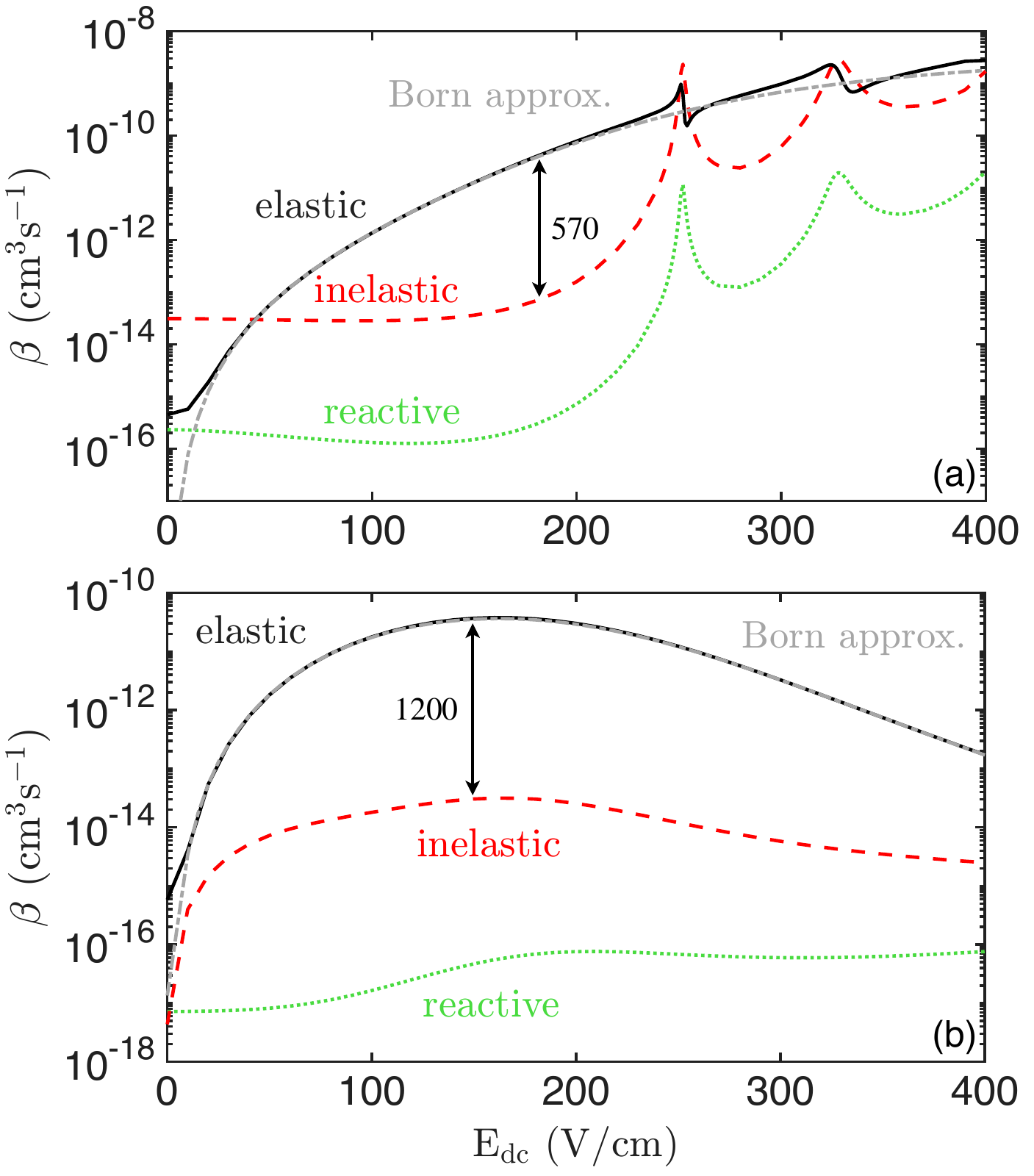}
    \caption{ Elastic (solid black), inelastic (dashed red) and reactive (dotted green) rate coefficients as a function of applied electric field for colliding molecules prepared in the $| \tilde{J}, m_J; \tilde{\epsilon}_K \rangle = | 3/2, +1/2; + \rangle$ state (upper panel) and the $| \tilde{J}, m_J; \tilde{\epsilon}_K \rangle = | 1/2, +1/2; + \rangle$ state (lower panel).  }
    \label{fig:ICS_vs_Edc__mJ1half}
\end{figure}

We find that the rate coefficients obtained through Eq.~(\ref{eq:Born_approximation}), plotted as dashed-dotted gray curves, compare remarkably well with the elastic integral cross sections from full numerical close-coupling calculations (solid black curves) in figures \ref{fig:ICS_vs_Edc__JmJ3half} and \ref{fig:ICS_vs_Edc__mJ1half}. 
These curves only deviate significantly near zero field and in the presence of shape resonances, the latter resulting from the emergence of octatomic field-linked (FL) bound states \cite{Avdeenkov03_PRL}--weakly bound [CaNH$_2$]$_2$ dimer states resulting from long-range dipolar attraction and short-range van der Waals repulsion.  
Subplots (a) and (b) in Fig.~\ref{fig:ICS_vs_Edc__JmJ3half} plot the adiabatic potentials that asymptotically connect to $| \tilde{J}, m_J; \tilde{\epsilon}_K \rangle = | 3/2, +3/2; + \rangle$, at electric field values below (${\rm E}_{\rm dc} = 80$ V/cm) and above (${\rm E}_{\rm dc} = 160$ V/cm) a FL resonance respectively.      
The adiabatic curves are generated with $L = 1,3,5$ partial waves, restricted to the conserved ${\cal M} = m_{J_1} + m_{J_2} + m_L = 3$ subspace with subscripts on $J$ indexing molecules $1$ and $2$, while $m_L$ is the $L$-projection quantum number. 
The lowest partial wave adiabat in each subplot, highlighted in red and thickened, indicates that the purely repulsive potential at low fields develops a long-range potential well as ${\rm E}_{\rm dc}$ increases. 
These wells can host several FL bound states, with each new bound state corresponding to a resonance feature as seen in the lower panel subplot of Fig.~\ref{fig:ICS_vs_Edc__JmJ3half}.

Similar FL resonances also occur for molecules prepared in $| \tilde{J}, m_J; \tilde{\epsilon}_K \rangle = | 3/2, +1/2; + \rangle$ observed in the upper panel of Fig.~\ref{fig:ICS_vs_Edc__mJ1half}, with elastic scattering rates varying by more than six orders of magnitude over the plotted field range.   
Recently observed between ultracold diatomic molecules \cite{Chen23_Nature, Chen24_Nature}, FL bound states have also been predicted between polyatomic CaOH \cite{Augustovicova19_NJP}, although these hexatomic molecules have yet to be experimentally demonstrated. 
Our prediction of octatomic FL states therefore extends this phenomenon to even larger dipolar molecules, indicating that this pairing mechanism could be generic to sufficiently dipolar molecules with a doublet structure.

\subsection{ Collisional loss and prospects for evaporation }

Turning to the loss cross sections, we find that over the range of electric fields studied here, reactive loss remains orders of magnitude smaller than the inelastic loss due to the high short-range repulsive potential barriers present in the intermolecular potential.    
At fields of ${\rm E}_{\rm} < 200$ V/cm below what is necessary to support FL bound states, we find that preparing the molecules in $| \tilde{J}, m_J; \tilde{\epsilon}_K \rangle = | 3/2, +1/2; + \rangle$ or $| 1/2, +1/2; + \rangle$ can result in an elastic-to-loss rate ratio of $\approx 570$ or $\approx 1200$ respectively, shown in Fig.~\ref{fig:ICS_vs_Edc__mJ1half}. 
We note that unlike collisional shielding in diatomic molecules, application of the electric field does not actually suppress collisional loss between CaNH$_2$ colliders, but rather enhances elastic collisions to exceed the former.

\begin{figure}[ht]
    \centering
    \includegraphics[width=\linewidth]{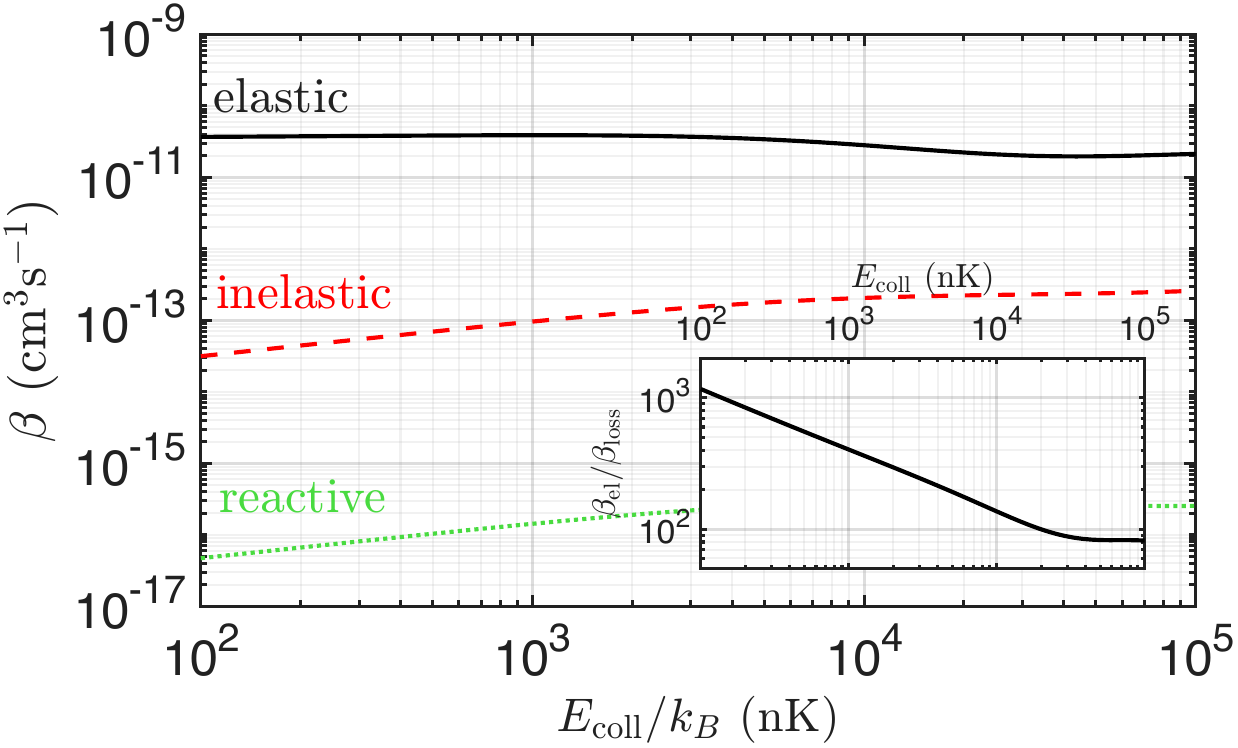}
    \caption{ Collisional elastic (solid black), inelastic (dashed red) and reactive (dotted green) rate coefficients of CaNH$_2$ prepared in $| \tilde{J}, m_J; \tilde{\epsilon}_K \rangle = | 1/2, +1/2; + \rangle$ and subject to ${\rm E}_{\rm dc} = 150$ V/cm, as a function of collision energy. The inset plots the ratio of elastic-to-loss rate coefficients as a function of collision energy. }
    \label{fig:fermionicICS(JmJ1half)_vs_Ecoll}
\end{figure}

The calculations done thus far assume a collision energy of $E_{\rm coll}/k_B = 100$ nK, but we observe that the elastic rate coefficient remains over eighty times above the loss rate coefficient even at collision energies of 100 $\mu$K, with the latter remaining at the $10^{-13}$ cm$^3$s$^{-1}$ level as showcased in Fig.~\ref{fig:fermionicICS(JmJ1half)_vs_Ecoll}. These rate coefficients are obtained with the molecules prepared in $| \tilde{J}, m_J; \tilde{\epsilon}_K \rangle = | 1/2, +1/2; + \rangle$ and subject to ${\rm E}_{\rm dc} = 150$ V/cm. At this field value, the induced dipole moment is $0.25$ D. 
Lower elastic-to-loss rate ratios have proven sufficient to allow efficient evaporative cooling of fermionic KRb molecules \cite{Li21_Nat, Lin26_arxiv}. 
Our results thus promise that evaporative cooling of CaNH$_2$ could commence directly after laser cooling to sub-milliKelvin temperatures, providing a route to realizing deeply degenerate Fermi gases of polyatomic molecules.

The main qualitative results for CaNH$_2$ of highly elastic collisions and octatomic FL states are also observed with SrNH$_2$, included in App.~\ref{app:SrNH2}. Other heavier variants of alkaline-earth monoamides such as BaNH$_2$ or RaNH$_2$ that are relevant to electron electric dipole moment measurements, are thus also expected to exhibit similar favorable collisional properties, promising a general pathway to quantum degenerate samples of this class of polyatomic molecules.

\section{ Summary and Outlook \label{sec:summary} }

We studied the ultracold collisions of asymmetric top CaNH$_2$ molecules in the presence of a static electric field. 
Leveraging the modest zero-field $K$-doublet splitting in the ground vibronic $N=1$ manifold, electric fields below 1 kV/cm induce large lab-frame dipole moments while preserving repulsive rotational van der Waals barriers against short-range loss. 
Our close-coupling scattering calculations produce elastic-to-loss rate ratios exceeding three orders of magnitude at ultracold temperatures, with high elasticity persisting up to collision energies of 100 $\mu$K. Furthermore, increasing the electric field in select molecular states induces weakly bound octatomic field-linked dimers of [CaNH$_2$]$_2$, extending the field-linked pairing mechanism to polyatomic asymmetric tops. 
We expect these collisional features to be generic across dipolar alkaline-earth monoamides, as demonstrated through the analogous highly elastic collisions and field-linked resonances in SrNH$_2$.

The calculated ratios of elastic and loss rate coefficients are favorable for evaporative cooling to quantum degenerate samples of polyatomic molecules.     
Such prospects for achieving low entropy ensembles of ground state asymmetric tops open opportunities for utilizing the long-lived $K$-doublet for quantum simulation, quantum computation, and precision searches of beyond the Standard Model physics.

There are several natural extensions to this work.  
For one, the application of magnetic fields can be used to tune the spin-rotation fine structure of the molecules. The effect of this tuning on the ultracold collision dynamics has yet to be explored.
Additionally, microwave fields could be used to further dress the interactions, possibly allowing loss rate suppression or field-linked state tunability by mixing in other rotational manifolds \cite{Wang26_arxiv, Ho26_PRR, Wang26_arxiv2}. 
Controlled variation of these microwaves or the static electric field could enable electroassociation of the octatomic field-linked states \cite{Quemener23_PRL}.   
We leave such investigations to a future work.

\begin{acknowledgments}

We thank J. L. Bohn for valuable discussions, the Doyle CaNH$_2$ team for useful experimental insights, and H. R. Sadeghpour for a thorough reading of this work.  
Support for this project was provided by ITAMP with funding from the National Science Foundation. 

\end{acknowledgments}

\begin{appendix}

\section{ Collision rates for \texorpdfstring{SrNH$_2$}{SrNH2} \label{app:SrNH2} }

\begin{figure}[ht]
    \centering
    \includegraphics[width=1\linewidth]{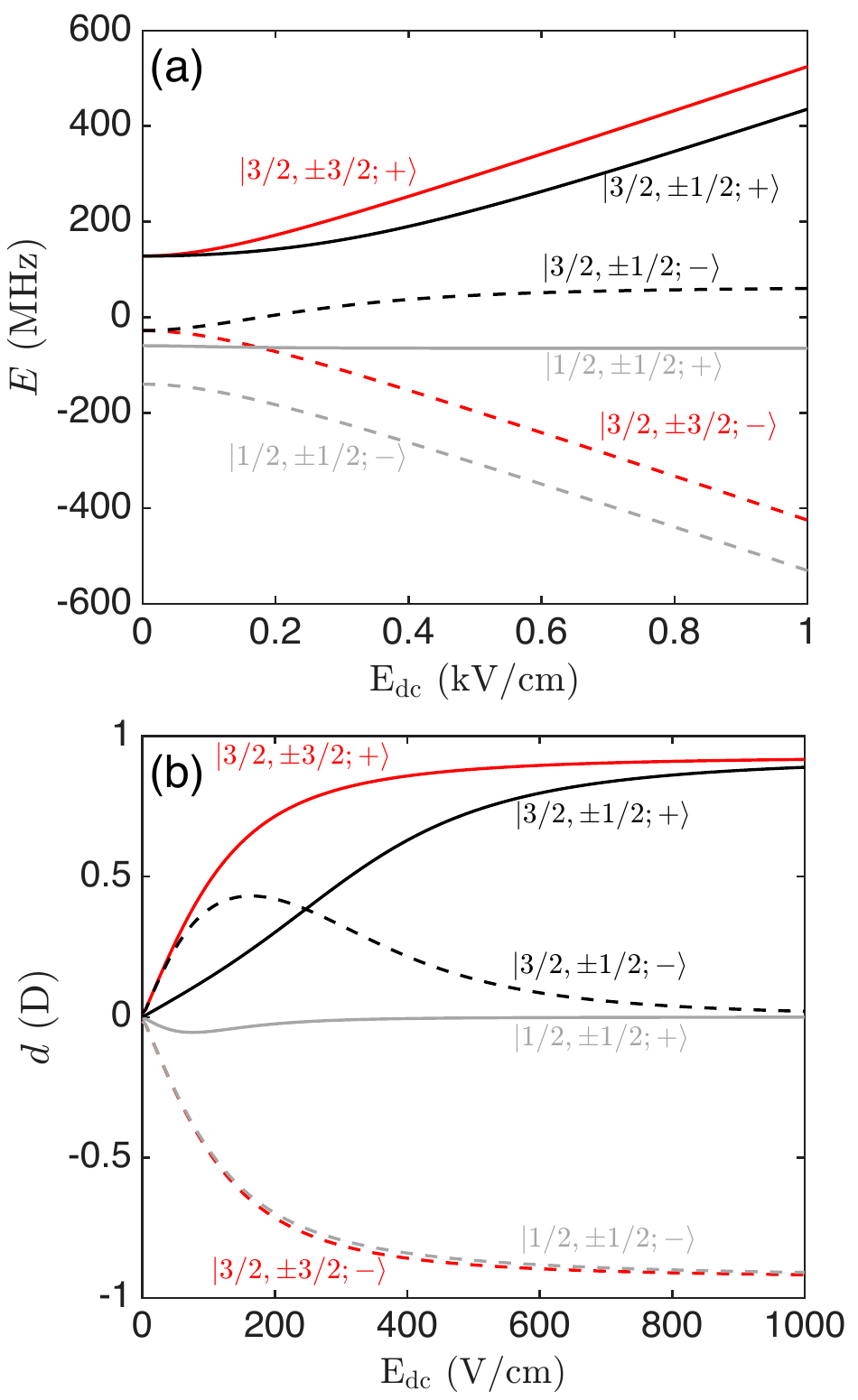}
    \caption{ (a) The molecular spectrum of SrNH$_2$ in the reduced $N=1, |K|=1$ subspace, as a function of electric field. (b) Corresponding induced dipole moments as a function of electric field. The state labels and colors follow those in Fig.~\ref{fig:CaNH2_rotationsNdipoles}. }
    \label{fig:SrNH2_rotationsNdipoles}
\end{figure}

In this appendix section, we perform close-coupling calculations for the ultracold collisions between SrNH$_2$ molecules. 
The molecular parameters for SrNH$_2$ are provided in Tab.~\ref{tab:SrNH2_constants}, showcasing a smaller $K$-doublet splitting of $130$ MHz alongside a larger dipole moment compared to CaNH$_2$.
A consequence of these parameters is that not all $\epsilon_K = +$ states are above all $\epsilon_K = -$ states in SrNH$_2$, seen in Fig.~\ref{fig:SrNH2_rotationsNdipoles}. 
Therefore, we do not expect a similarly high ratio of elastic-to-loss rates for $|\tilde{J}, m_J; \tilde{\epsilon}_K\rangle = |1/2, 1/2, +\rangle$ in SrNH$_2$ as in CaNH$_2$. As such, we focus analysis here on molecules prepared in $|\tilde{J}, m_J; \tilde{\epsilon}_K\rangle = |3/2, 3/2, +\rangle$. 

\begin{table}[ht]
    \centering
    \begin{tabular}{lcc}
        \hline
        constant & symbol & value \\
        \hline
        dipole moment \cite{Yang26_arxiv} & $d_0$ & 1.86 D \\
        rotational constants \cite{Thompsen00_CPL} & $A$ & 394.340 GHz \\
        & $B$ & 6.7902961 GHz \\
        & $C$ & 6.6595159 GHz \\
        spin-rotation couplings \cite{Thompsen00_CPL} & $\epsilon_{xx}$ & 160.4 MHz \\
        & $\epsilon_{yy}$ & 59.740 MHz \\
        & $\epsilon_{zz}$ & 89.657 MHz \\
        \hline
    \end{tabular}
    \caption{ Molecular constants for SrNH$_2$. }
    \label{tab:SrNH2_constants}
\end{table}

\begin{figure}[ht]
    \centering
    \includegraphics[width=\linewidth]{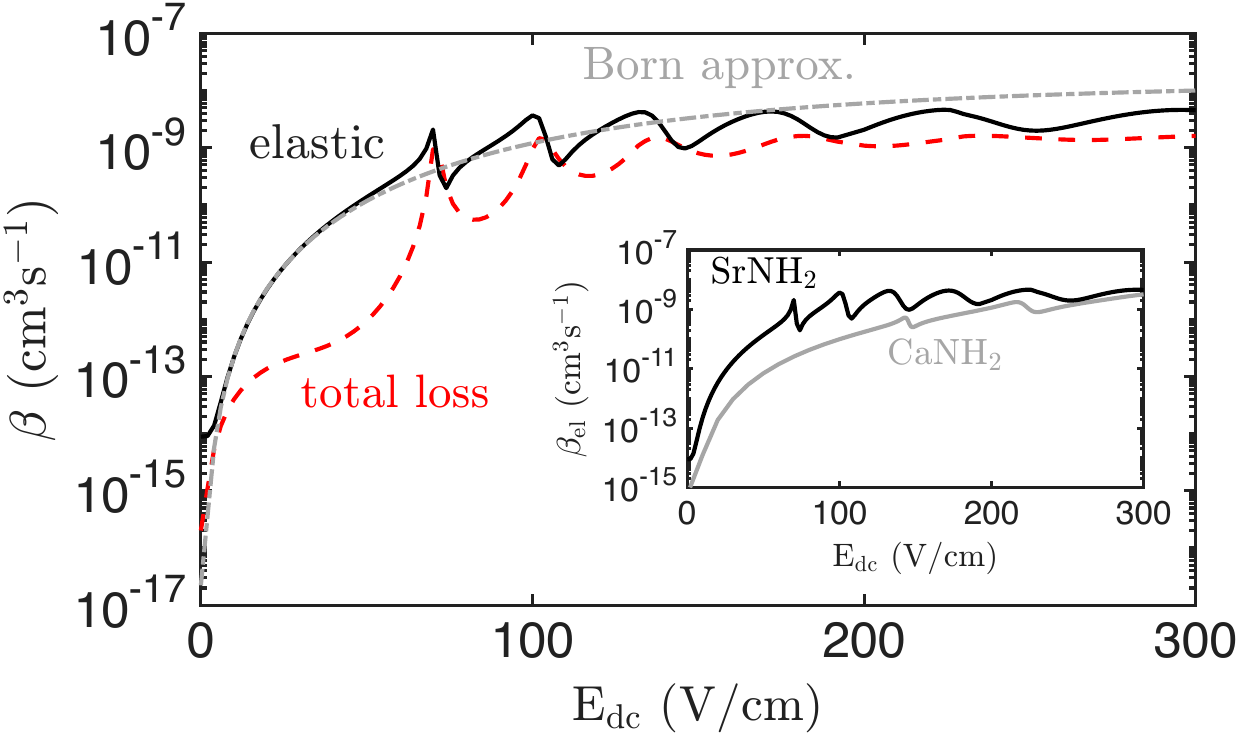}
    \caption{ Elastic (solid black) and total loss (dashed red) rate coefficients between ultracold SrNH$_2$ molecules as a function of applied electric field. The molecules are prepared in the $| \tilde{J}, m_J; \tilde{\epsilon}_K \rangle = | 3/2, +3/2; + \rangle$ state. }
    \label{fig:SrNH2_ICS(JmJ3half)_vs_Edc}
\end{figure}

From close-coupling calculations, we find that an elastic-to-loss ratio of $\approx 150$ is achieved at ${\rm E}_{\rm dc} = 45$ V/cm with a loss rate coefficient of $\approx 5.5 \times 10^{-13}$ cm$^{3}$s$^{-1}$, seen in Fig.~\ref{fig:SrNH2_ICS(JmJ3half)_vs_Edc}. 
The figure plots the elastic (solid black) and total loss (dashed red) rate coefficients, the latter comprising the sum of inelastic and reactive rate coefficients.   
For larger electric fields, several FL resonances are also observed that indicate the existence of weakly bound octatomic [SrNH$_2$]$_2$ dimers. 
The inset compares the elastic rate coefficient $\beta_{\rm el}$ between CaNH$_2$ (solid black) and SrNH$_2$ (solid gray) as a function of electric field, showcasing a more rapid increase in $\beta_{\rm el}$ for SrNH$_2$ due to its smaller $K$-doublet.

A broader survey of field parameters and molecular states could be performed to find more favorable conditions for evaporative cooling of SrNH$_2$ to quantum degeneracy. Such analysis is deferred to a future, more focused work. 

\end{appendix}


\nocite{*}
\bibliography{main.bib} 
\vspace{\fill}

\include{SI.tex}
\end{document}

%% file: SI.tex
\onecolumngrid
 


\newpage
\renewcommand{\thefigure}{S\arabic{figure}}

\setcounter{figure}{0}
\setcounter{section}{0}
\setcounter{subsection}{0}

\begin{center}

{\large\bfseries Supplementary Information\par}

\vspace{0.5em}

{\large\bfseries
Highly elastic collisions of ultracold asymmetric top molecules in a static electric field
\par}

\vspace{0.8em}

Reuben R. W. Wang$^{1,2}$

\vspace{0.5em}

{\small
$^{1}$\textit{ITAMP, Center for Astrophysics $|$ Harvard \& Smithsonian, Cambridge, Massachusetts 02138, USA}

$^{2}$\textit{Department of Physics, Harvard University, Cambridge, Massachusetts 02138, USA}
\par}

\end{center}

\section{ Hamiltonian matrix elements }

For the nearly prolate symmetry of CaNH$_2$, the matrix elements of the molecular Hamiltonian in the symmetric top basis are computed as 
\begin{align}
    & \bra{ N', K', M' }
    \mathcal{H}_{\rm rot}
    \ket{ N, K, M } \nonumber\\
    &=
    \bra{ N', K', M' }
    \left(
    B N_x^2 + C N_y^2 + A N_z^2
    \right)
    \ket{ N, K, M } \nonumber\\
    &=
    \left[
    \frac{1}{2} (B + C) N (N + 1) 
    +
    \frac{1}{2} \left( 2 A - (B + C) \right) K^2
    \right] \delta_{N',N} \delta_{K',K} \delta_{M',M} \nonumber\\
    &\quad 
    +
    \frac{1}{4} (B - C)
    \delta_{N',N}
    \delta_{K',K+2}
    \delta_{M',M}
    \sqrt{ N (N + 1) - K (K + 1) }
    \sqrt{ N (N + 1) - (K + 1) (K + 2) }
    \nonumber\\
    &\quad 
    +
    \frac{1}{4} (B - C)
    \delta_{N',N}
    \delta_{K',K-2}
    \delta_{M',M}
    \sqrt{ N (N + 1) - K (K - 1) }
    \sqrt{ N (N + 1) - (K - 1) (K - 2) },
\end{align} 
where the degeneracies in $K$ are lifted due to the broken cylindrical symmetry along the molecules axis.
Notably, $N$ and $M$ remain good quantum numbers in the basis that diagonalizes the asymmetric top Hamiltonian.  
To compute the matrix elements of the spin-rotation interaction, it is more appropriate to adopt the couple basis with total angular momentum $\boldsymbol{J} = \boldsymbol{N} + \boldsymbol{S}$, and eigenstates $|N,S,J; m_J, K\rangle$. In this basis, the matrix elements are
\begin{align}
    & \langle N',S,J'; m'_J, K' |
    {\cal H}_{\rm sr}
    | N,S,J; m_J, K \rangle \nonumber\\
    &=
    \sum_{\alpha} \epsilon_{\alpha \alpha} 
    \langle N',S,J'; m'_J, K' |
    N_{\alpha} S_{\alpha}
    | N,S,J; m_J, K \rangle \nonumber\\
    &= 
    \langle N',S,J'; m'_J, K' |
    \bigg[
    \frac{ 1 }{ 4 }
    (\epsilon_{xx} - \epsilon_{yy}) 
    (N_+ S_+ + N_- S_-)
    +
    \frac{1}{4}
    (\epsilon_{xx} + \epsilon_{yy}) 
    (N_+ S_- + N_- S_+) \nonumber\\
    &\quad\quad\quad\quad
    \quad\quad\quad\quad
    \quad\quad\quad\quad
    \quad\quad\quad\quad
    \quad\quad\quad\quad
    \quad\quad\quad\quad
    +
    \epsilon_{z z} N_{z} S_{z}
    \bigg]
    | N,S,J; m_J, K \rangle \nonumber\\
    &=
    f(N,J)
    \left[
    \epsilon_{z z} K^2
    +
    \frac{1}{2}
    (\epsilon_{xx} + \epsilon_{yy}) 
    ( N (N + 1) - K^2 )
    \right]
    \delta_{N',N}
    \delta_{J',J}
    \delta_{K',K}
    \delta_{m'_J,m_J} \nonumber\\
    &\quad 
    +
    \frac{1}{4}
    (\epsilon_{xx} - \epsilon_{yy}) 
    f(N,J)
    \sqrt{ (N \mp K) (N \mp K - 1) (N \pm K + 1) (N \pm K + 2) }
    \delta_{N',N}
    \delta_{J',J}
    \delta_{K',K \pm 2}
    \delta_{m'_J,m_J} \nonumber\\
    &\quad 
    +
    \delta_{N',N \pm 1} 
    ( \ldots ),
\end{align}
where $f(N,J) = [J (J + 1) - N (N + 1) - S (S + 1)] / [2 N (N + 1)]$.
We only consider the terms that preserve the $N$ quantum number as we will restrict all considerations to a single $N$ manifold as argued below. 
Conveniently, the matrix elements of ${\cal H}_{\rm rot}$ remain the same in the coupled $J$ basis since there are only off-diagonal terms in $K$, which is a well-defined quantum number in both bases. All other quantum numbers are conserved by ${\cal H}_{\rm rot}$.

\subsection{ Restriction to the \textit{N}=1 manifold }

To simplify our discussions, we first restrict ourselves to the $N = 1, |K| = 1$ manifold ($J = 1/2, 3/2$). 
This truncation is warranted as we notice that the coefficients multiplying the $N(N+1)$ and $K^2$ symmetric top terms in ${\cal H}_{\rm mol}$ all lie in the 10s to 100s of GHz range. 
Meanwhile, the $K$-doubling and spin-rotation structure lie in a 10s to 100s of MHz range, establishing a large separation of energy scales for projection into this closely degenerate subspace.  
We will consider the application of electric fields that are not large enough to mix in other rotational manifolds appreciably.  
In this reduced subspace, the bare molecular Hamiltonian reduces to 
\begin{align}
    & \bra{ J'; m'_J, K' }
    \mathcal{H}_{\rm rot}
    \ket{ J; m_J, K } \nonumber\\
    &=
    \sum_{N', N}
    \sum_{M', M}
    \sum_{m'_S, m_S}
    \langle J'; m'_J, K' |
    N'; m'_S, M', K' \rangle 
    \langle N'; m'_S, M', K' |
    {\cal H}_{\rm rot}
    | N; m_S, M, K \rangle \nonumber\\
    &\quad\quad\quad\quad
    \quad\quad\quad\quad \times 
    \langle N; m_S, M, K
    | J; m_J, K \rangle \nonumber\\
    &=
    \sum_{N, M}
    \sum_{m'_S, m_S}
    \langle J'; m'_J, K' |
    N; m'_S, M, K' \rangle 
    \langle N; m'_S, M, K' |
    {\cal H}_{\rm rot}
    | N; m_S, M, K \rangle  
    \langle N; m_S, M, K
    | J; m_J, K \rangle \nonumber\\
    &=
    \sum_{N, M}
    \sum_{m_S}
    \langle J'; m'_J, K' |
    N; m_S, M, K' \rangle 
    \langle N; m_S, M, K' |
    {\cal H}_{\rm rot}
    | N; m_S, M, K \rangle  
    \langle N; m_S, M, K
    | J; m_J, K \rangle \nonumber\\
    &=
    \sum_{M}
    \langle J'; m_J, K' |
    1; m_J-M, M, K' \rangle 
    \langle 1; m_J-M, M, K
    | J; m_J, K \rangle 
    \delta_{m'_J, m_J} \nonumber\\
    &\quad\quad\quad\quad \times 
    \langle 1; m_J-M, M, K' |
    {\cal H}_{\rm rot}
    | 1; m_J-M, M, K \rangle \nonumber\\
    &= 
    \delta_{m'_J, m_J}
    \sum_{M}
    \langle J'; m_J, K' |
    1; m_J-M, M, K' \rangle 
    \langle 1; m_J-M, M, K
    | J; m_J, K \rangle \nonumber\\
    &\quad\quad\quad\quad \times 
    \left[
    \left( A + \frac{B + C}{2} \right)
    \delta_{K',K} 
    +
    \frac{1}{4} (B - C)
    \sqrt{ 2 - K (K \pm 1) }
    \sqrt{ 2 - (K \pm 1) (K \pm 2) }
    \delta_{K',K \pm 2}
    \right] \nonumber\\
    &= 
    \delta_{m'_J, m_J}
    \left[
    \left( A + \frac{B + C}{2} \right)
    \delta_{K',K} 
    +
    \frac{1}{4} (B - C)
    \sqrt{ 2 - K (K \pm 1) }
    \sqrt{ 2 - (K \pm 1) (K \pm 2) }
    \delta_{K',K \pm 2}
    \right] \nonumber\\
    &\quad\quad\quad\quad \times 
    \sum_{M}
    \langle J'; m_J |
    1; m_J-M, M \rangle 
    \langle 1; m_J-M, M
    | J; m_J \rangle.
\end{align}
With Clebsch-Gordan coefficients
\begin{align}
    \langle N, M; S, m_S | J, m_J \rangle 
    &=
    (-1)^{-N+S-m_J}
    \sqrt{ 2 J + 1 }
    \begin{pmatrix}
        N & S & J \\
        M & m_S & -m_J
    \end{pmatrix},
\end{align}
we see that
\begin{align}
    & \sum_{M=-1}^{+1}
    \langle J'; m_J |
    1; m_J-M, M \rangle 
    \langle 1; m_J-M, M
    | J; m_J \rangle \nonumber\\
    &=
    \sqrt{ (2 J' + 1) (2 J + 1) }
    \sum_{M=-1}^{+1}
    \begin{pmatrix}
        1 & 1/2 & J' \\
        M & m_J - M & -m_J
    \end{pmatrix}
    \begin{pmatrix}
        1 & 1/2 & J \\
        M & m_J - M & -m_J
    \end{pmatrix} 
    =
    \delta_{J', J},
\end{align}
which gives us that
\begin{align}
    & \bra{ J'; m'_J, K' }
    \mathcal{H}_{\rm rot}
    \ket{ J; m_J, K } \nonumber\\
    &\approx 
    \left[
    \left( A + \frac{B + C}{2} \right)
    \delta_{K',K}
    +
    \frac{1}{4} (B - C)
    \sqrt{ 2 - K (K \pm 1) }
    \sqrt{ 2 - (K \pm 1) (K \pm 2) }
    \delta_{K',K \pm 2}
    \right] 
    \delta_{J',J} 
    \delta_{m'_J, m_J}. \nonumber
\end{align}

Then considering the spin-rotation coupling term with $N = 1$ and $S = 1/2$, it too is reduced to
\begin{align}
    & \langle J'; m'_J, K' |
    {\cal H}_{\rm sr}
    | J; m_J, K \rangle \nonumber\\
    &=
    f_1(J)
    \left[
    \epsilon_{z z} K^2
    +
    \frac{1}{2}
    (\epsilon_{xx} + \epsilon_{yy}) 
    ( 2 - K^2 )
    \right]
    \delta_{J',J}
    \delta_{K',K}
    \delta_{m'_J,m_J} \nonumber\\
    &\quad 
    +
    \frac{1}{4}
    (\epsilon_{xx} - \epsilon_{yy}) 
    f_1(J)
    \sqrt{ (1 \mp K) (\mp K) (2 \pm K) (3 \pm K) }
    \delta_{J',J}
    \delta_{K',K \pm 2}
    \delta_{m'_J,m_J},
\end{align}
where $f_1(J) = [J (J + 1)/4 - 11/16]$. Ignoring the constant diagonal energy offset, the total bare molecular Hamiltonian is given by
\begin{align}
    \bra{ J'; m'_J, K' }
    \mathcal{H}
    \ket{ J; m_J, K }
    &=
    f_1(J)
    \left[
    \epsilon_{z z}
    +
    \frac{1}{2}
    (\epsilon_{xx} + \epsilon_{yy}) 
    \right]
    \delta_{J',J}
    \delta_{K',K}
    \delta_{m'_J,m_J} \\
    &\quad 
    +
    \frac{1}{4} 
    \Big[
    (B - C)
    \sqrt{ 2 - K (K \pm 1) }
    \sqrt{ 2 - (K \pm 1) (K \pm 2) } \nonumber\\
    &\quad\quad\:\: 
    +
    f_1(J)
    (\epsilon_{xx} - \epsilon_{yy}) 
    \sqrt{ (1 \mp K) (\mp K) (2 \pm K) (3 \pm K) }
    \Big]
    \delta_{J',J}
    \delta_{K',K \pm 2}
    \delta_{m'_J,m_J}. \nonumber
\end{align}

\subsection{ K-doublets }

The degeneracy in the eigenstates of $K = \pm 1$ in the pure prolate limit ($B = C$), result in non-degenerate parity eigenstates of the asymmetric top--odd and even parity combinations of the $\pm |K|$ states:
\begin{align} 
    \ket{ J; m_J, \varepsilon|K| }
    &=
    \frac{ 1 }{ \sqrt{ 2 } }
    \left(
    \ket{ J; m_J, +|K| }
    +
    \varepsilon
    \ket{ J; m_J, -|K| }
    \right),
\end{align}
where $\varepsilon = \pm 1$ denotes the state parity. 
In the reduced $N = 1, |K| = 1$ subspace, we can construct a simple $2 \times 2$ matrix representation of the molecular Hamiltonian for given $J$ and $m_J$ values in the basis $| J; m_J, K \rangle = | J; m_J, -1 \rangle, | J; m_J, +1 \rangle$:
\begin{align}
    \boldsymbol{{\cal H}}_{\rm mol}^{(J, m_J)}
    &=
    \begin{pmatrix}
        E_0(J) & \delta{E}(J) \\
        \delta{E}(J) & E_0(J)
    \end{pmatrix}
\end{align}
where $E_0(J) = f_1(J) [ \epsilon_{zz} + (\epsilon_{xx} + \epsilon_{yy})/2 ]$ and $\delta{E}(J) = [(B - C) + f_1(J) (\epsilon_{xx} - \epsilon_{yy})]/2$. Diagonalizing this matrix immediately gives the eigenstates in Eq.~(\ref{eq:parity_eigenstates}) with corresponding eigenenergies $E_{\varepsilon} = \varepsilon \delta{E}$:
\begin{subequations} \label{eq:Kdoublet_basis}
\begin{align}
    \ket{ J; m_J, \varepsilon=-1 }
    &=
    \frac{ 1 }{ \sqrt{ 2 } }
    \left(
    \ket{ J; m_J, K=+1 }
    -
    \ket{ J; m_J, K=-1 }
    \right): 
    \quad 
    E_{-} 
    =
    E_0 - \delta{E}, \\
    \ket{ J; m_J, \varepsilon=+1 }
    &=
    \frac{ 1 }{ \sqrt{ 2 } }
    \left(
    \ket{ J; m_J, K=+1 }
    +
    \ket{ J; m_J, K=-1 }
    \right): 
    \quad 
    E_{+} 
    =
    E_0 + \delta{E},
\end{align}
\end{subequations}
opposite parity states referred to as $K$-doublets. 

\subsection{ Application of a static electric field }

The dipole-field coupling then has matrix elements given in the uncoupled basis by
\begin{align}
    \langle N', K', M'|
    -\boldsymbol{d}_0 \cdot \boldsymbol{{\rm E}}_{\rm dc}
    | N, K, M \rangle
    &=
    -d_0 {\rm E}_{\rm dc}
    (-1)^{M' - K'}
    \sqrt{ (2 N' + 1) (2 N + 1) } \nonumber\\
    &\quad\quad \times 
    \begin{pmatrix}
        N' & 1 & N \\
        -K' & 0 & K 
    \end{pmatrix}
    \begin{pmatrix}
        N' & 1 & N \\
        -M' & 0 & M 
    \end{pmatrix}.
\end{align}
In the reduced $N=1,|K|=1$ subspace, the matrix elements in the coupled basis representation is given by
\begin{align}
    & \langle J'; m'_J, K'|
    -\boldsymbol{d}_0 \cdot \boldsymbol{{\rm E}}_{\rm dc}
    | J; m_J, K \rangle \nonumber\\
    &=
    \sum_{M', M}
    \sum_{m'_S, m_S}
    \langle J'; m'_J, K'
    | K', M', m'_S \rangle \nonumber\\
    &\quad\quad\quad\quad \times 
    \langle K', M', m'_S|
    -\boldsymbol{d} \cdot \boldsymbol{{\rm E}}_{\rm dc}
    | K, M, m_S \rangle\langle K, M, m_S |
    J; m_J, K \rangle \nonumber\\
    &=
    -d_0 {\rm E}_{\rm dc}
    \sum_{M', M}
    \sum_{m'_S, m_S}
    \langle J'; m'_J, K'
    | K', M', m'_S \rangle\langle K, M, m_S |
    J; m_J, K \rangle \nonumber\\
    &\quad\quad\quad\quad \times  
    3 (-1)^{M' - K'}
    \begin{pmatrix}
        1 & 1 & 1 \\
        -K' & 0 & K 
    \end{pmatrix}
    \begin{pmatrix}
        1 & 1 & 1 \\
        -M' & 0 & M 
    \end{pmatrix}
    \delta_{m'_S, m_S} \nonumber\\
    &=
    -d_0 {\rm E}_{\rm dc}
    \sum_{M, m_S}
    \langle J'; m'_J, K
    | K, M, m_S \rangle\langle K, M, m_S |
    J; m_J, K \rangle \nonumber\\
    &\quad\quad\quad\quad \times  
    3 (-1)^{M - K}
    \begin{pmatrix}
        1 & 1 & 1 \\
        -K & 0 & K 
    \end{pmatrix}
    \begin{pmatrix}
        1 & 1 & 1 \\
        -M & 0 & M 
    \end{pmatrix} \nonumber\\
    &=
    -\frac{ 1 }{ 2 }
    d_0 {\rm E}_{\rm dc} 
    \delta_{K', K}
    \sum_{M, m_S}
    \langle J'; m'_J, K
    | K, M, m_S \rangle
    ( K M )
    \langle K, M, m_S |
    J; m_J, K \rangle,
\end{align}
Because $m_J = M + m_S$, we see from the Clebsch-Gordan coefficients that $m_J$ must be conserved by the electric field, so we have that 
\begin{align}
    & \langle J'; m'_J, K'|
    -\boldsymbol{d}_0 \cdot \boldsymbol{{\rm E}}_{\rm dc}
    | J; m_J, K \rangle \nonumber\\
    &=
    -\frac{ 1 }{ 2 }
    d_0 {\rm E}_{\rm dc}
    \delta_{m'_J, m_J}
    \delta_{K', K}
    \sum_{M, m_S}
    \langle J'; m_J, K
    | K, M, m_S \rangle
    ( K M )
    \langle K, M, m_S |
    J; m_J, K \rangle.
\end{align}
The static electric field will couple states of different total angular momentum $J$ and parity $\epsilon$, resulting in the field-dressed eigenstates $|\tilde{J}; m_J, \tilde{\epsilon}\rangle$ that adiabatically correlate the zero-field $|{J}; m_J, {\epsilon}\rangle$ states.


\section{ Long-range dipole-dipole interactions }

We take that the two molecules interact predominantly at long-range via dipole-dipole interactions:
\begin{align}
    V_{\rm dd}
    &=
    -\frac{ \sqrt{ 6 } }{ 4 \pi \epsilon_0 r^3 }
    \sum_{p=-2}^2
    (-1)^p C_{2, -p}(\theta, \phi)
[ \boldsymbol{d}_{0,A} \otimes \boldsymbol{d}_{0,B} ]_p^{(2)} \nonumber\\
    &=
    -\frac{ \sqrt{ 6 } }{ 4 \pi \epsilon_0 r^3 }
    \sum_{p=-2}^2
    (-1)^p C_{2, -p}(\theta, \phi)
    \sum_{p'=-1}^1
    \bra{ 1, p'; 1, p - p' }\ket{ 2, p }
    d_{A}^{(p')} d_{B}^{(p - p')},
\end{align}
where $C_{\ell, m}(\theta, \phi) = Y_{\ell, m}(\theta, \phi) \sqrt{ 4\pi/(2\ell + 1) }$ are reduced spherical harmonics, $\bra{ \ell_1, m_1; \ell_2, m_2 }\ket{ \ell_3, m_3 }$ are Clebsch-Gordan coefficients, while $d_0^{(0)} = d_Z$ and $d_0^{(\pm 1)} = \mp ( d_X \pm i d_Y ) / \sqrt{2}$ are the rank-1 spherical dipole tensors. 
The dipole-dipole interaction matrix elements are given in the pair symmetric top basis as
\begin{align}
    & \bra{ N'_A, K'_A, M'_A; N'_B, K'_B, M'_B } V_{\rm dd} \ket{ N_A, K_A, M_A;  N_B, K_B, M_B } \nonumber\\
    =&
    -\frac{ \sqrt{ 6 } }{ 4 \pi \epsilon_0 r^3 }
    \sum_{p=-2}^2
    (-1)^p C_{2, -p}(\theta, \phi)
    \sum_{p'=-1}^1
    \bra{ 1, p'; 1, p - p' }\ket{ 2, p }
    \bra{ N'_A, K'_A, M'_A } d_{0,A}^{(p')} \ket{ N_A, K_A, M_A } \nonumber\\
    &\qquad\qquad\qquad\qquad
    \qquad\qquad\qquad\qquad 
    \qquad\qquad \times 
    \bra{ N'_B, K'_B, M'_B } d_{0,B}^{(p - p')} \ket{ N_B, K_B, M_B } \nonumber\\
    =&
    -\frac{ C_{2, -q}(\theta, \phi) }{ 4 \pi \epsilon_0 r^3 }
    \sqrt{ 30 }
    \begin{pmatrix}
        1 & 1 & 2 \\
        q_A & q_B & -q
    \end{pmatrix}
    \nonumber\\
    &\quad \times 
    d_0 (-1)^{M'_A - K'_A}
    \sqrt{ ( 2 N'_A + 1 ) ( 2 N_A + 1 ) }
    \begin{pmatrix}
        N'_A & 1 & N_A \\
        -K'_A & 0 & K_A
    \end{pmatrix}
    \begin{pmatrix}
        N'_A & 1 & N_A \\
        -M'_A & q_A & M_A
    \end{pmatrix}
    \nonumber\\
    &\quad \times 
    d_0 (-1)^{M'_B - K'_B}
    \sqrt{ ( 2 N'_B + 1 ) ( 2 N_B + 1 ) }
    \begin{pmatrix}
        N'_B & 1 & N_B \\
        -K'_B & 0 & K_B
    \end{pmatrix}
    \begin{pmatrix}
        N'_B & 1 & N_B \\
        -M'_B & q_B & M_B
    \end{pmatrix},
\end{align}
with the angular momentum transfer quantum numbers $q_A = M'_A - M_A$ and $q_B = M'_B - M_B$. 